\documentclass[12pt]{article}
\usepackage[T1]{fontenc}
\usepackage{calc}
\usepackage{setspace}
\usepackage{multicol}	% TODO: I don't need this package if the document is a single column one.
 
\begin{document}
\begin{center}
\textbf{{\large{}Fermion mass, a new interpretation and approach to fermionic vacuum and associated quantum field theories.}}
\end{center}

\begin{center}
\textbf{Jyoti Sekhar Bhattacharyya}
\end{center}

\begin{center}
\textbf{Dept. of Physics, Kanchrapara College,}
\end{center}

\begin{center}
\textbf{ Kanchrapara, North  24 -- Parganas, }
\end{center}

\begin{center}
\textbf{West Bengal, India }
\end{center}

\begin{center}

\end{center}

\textbf{\underline{{\large{}Abstract}}}

\begin{spacing}{1.24}
We argue that all fermionic states characterized by the (four) velocity not momentum of the fermion are always occupied by a single fermion. 
So there is no fermion number violation whatsoever. The mass of an elementary particle is not determined by field equations and is quite 
arbitrary. Though constant in the free state, interaction may cause a fermion to go from finite to zeromass state, interpreted as vacuum, 
and vice versa. We argue that a massless fermion can't be detected. Since occupation number is not fixed the arguments leading to the 
above conclusions cannot be extended to bosons. We calculate scattering amplitude for a typical process like Rutherford scattering in 
the light of the present discussion to show that there is no contradiction with the experimental results.
\end{spacing}

\begin{spacing}{1.24}
\newpage
       A recent work [1] in the field of neutrino oscillation has raised the issue of a new interpretation of fermionic vacuum, which we 
want to address in this paper. The preset work complements the earlier work [1].
\end{spacing}

\begin{spacing}{1.24}
We consider the Dirac equation
\end{spacing}

\begin{equation}
(i\gamma .\partial -m)\psi=0
\end{equation}

 to start with and expand $\psi$ in terms of  plane wave solutions

\begin{equation}
\psi= \sum_{\alpha, \vec k} [b_{\alpha \vec k}u_{\vec k}^{(\alpha)}e^{-ik . x} + d^\dagger_{\alpha \vec k} v_{\vec k}^{(\alpha)}
e^{ik . x}]
\end{equation}
Where 	the symbols have their usual meanings. We often replace integration by summation, Dirac $\delta$  by Kronecker $\delta$-function 
and ignore C-number factors, whenever considered not absolutely necessary.

One of the interesting aspects of equation (1) (or any other field equations for free fields) is, m the rest mass of the particle of 
the field $ \psi$  is quite arbitrary and can be changed at will under the scale transformation
   $m \rightarrow \alpha m$, $k \rightarrow \alpha k$. The (four) velocity of the particle $v = k/m$ remains fixed in this case. 
Though either of $\vec  v$ or $ \vec k$ can be treated as good quantum numbers for free stable particles (fixed m), such a scaling 
occurs in the case of neutrino oscillation [1], to make the theory selfconsistent. We therefore label free fermionic states corresponding 
to equation [2] with $\vec v$  instead of $\vec k$. Thus       

\begin{equation}
\psi= \sum_{\alpha, \vec v} [b_{\alpha \vec v}u_{\vec v}^{(\alpha)}e^{-imv.x} + d^\dagger_{\alpha \vec v} v_{\vec v}^{(\alpha)}
e^{imv.x}]
\end{equation}
 and the free Hamiltonian for the particles
\begin{eqnarray}
H_0&=& \int \psi^\dagger \partial_0 \psi d^3x\nonumber \\ &=& \sum_{\alpha,\vec v}b^\dagger_{\alpha\vec v} b_{\alpha \vec v}
\omega_{\vec v}
\end{eqnarray}
where $v_0 = (1+\vec v^2)^{1/2}$  and $\omega_{\vec v}=m(1+\vec v^2)^{1/2}$ is 
the on shell energy of the particle. From equation(1)  
\begin{equation}
(\gamma .v-1)u_{\vec v}^{(\alpha)} =0
\end{equation}
\begin{equation}
(\gamma .v+1)v_{\vec v}^{(\alpha)} =0
\end{equation}
So, 
\begin{equation}
u_{\vec v}^{(\alpha)}=\frac{(\gamma .v+1)}{2} u_0^{(\alpha)}
\end{equation}
This implies
\begin{eqnarray}
u_{\vec v}^{(1)}=\left ( \frac{1+v_0}{2}\right )^{1/2} \left (\begin{array}{c} 1\\ 0\\ v_z/(1+v_0)\\ v_+/(1+v_0) \end{array}\right )
\end{eqnarray}
\begin{eqnarray}
u_{\vec v}^{(2)}=\left ( \frac{1+v_0}{2}\right )^{1/2} \left (\begin{array}{c} 0\\ 1\\ v_-/(1+v_0)\\ -v_z/(1+v_0) \end{array}\right )
\end{eqnarray}
in the standard representation, where $v_\pm=v_x\pm iv_y$ and $\bar u_{\vec v}^{(\alpha)} u_{\vec v}^{(\alpha')}=\delta_{\alpha\alpha'}$
for on shell particles. A similar relation holds for the other spinor $v_{\vec v}^{(\alpha)}$.

From the equation (4) $b^\dagger_{\alpha \vec v} b_{\alpha \vec v}$ 
can be interpreted as the particle number for the state ($\alpha$, $\vec v$), which can take on values 0 and 1 if the following anti 
commutation relations hold.

\begin{equation}
\{	b_{\alpha \vec v}, b_{\alpha ' \vec v'}\} = 0
\{	b^\dagger_{\alpha \vec v}, b^\dagger_{\alpha ' \vec v'}\} = 0
\end{equation}
\begin{equation}
\{	b_{\alpha \vec v}, b^\dagger_{\alpha ' \vec v'}\} = \delta_{\alpha \alpha'}\delta_{\vec v \vec v'}
\end{equation}
	It is tempting to identify the system represented by equation (4) with quantum mechanical harmonic oscillators with 
lowering and raising operators $b_{\alpha,\vec v}$ and  $b^\dagger_{\alpha,\vec v}$. So we define the canonical coordinate and momenta

\begin{equation}
Q_{\alpha \vec v}=-i(b_{\alpha\vec v}-b^\dagger_{\alpha\vec v})/(2\omega_{\vec v})^{1/2}
\end{equation}
\begin{equation}
P_{\alpha \vec v}=(b_{\alpha \vec v}+b^\dagger_{\alpha \vec v})(\omega_{\vec v}/2)^{1/2}
\end{equation}

	From the canonical commutation relations
\begin{equation}
[Q_{\alpha\vec v},Q_{\alpha'\vec v'}]=[P_{\alpha\vec v},P_{\alpha'\vec v'}]=0
\end{equation}
\begin{equation}
[Q_{\alpha\vec v},P_{\alpha'\vec v'}]=i\delta_{\alpha\alpha'}\delta_{\vec v\vec v'}
\end{equation}

	We get,
\begin{equation}
[b_{\alpha\vec v},b_{\alpha'\vec v'}]=0
\end{equation}
\begin{equation}
[b_{\alpha\vec v},b^\dagger_{\alpha'\vec v'}]=-\delta_{\alpha\alpha'}\delta_{\vec v\vec v'}
\end{equation}

	The change of sign on the right hand side of equation (12) as compared to scalar fields, corresponds to the same of the commutator in equation (17). We would get the vacuous solution $b^\dagger_{\alpha\vec v}b_{\alpha\vec v}=0$ only,otherwise.

	From equations ( 11) and (17) we get,		
\begin{equation}
b^\dagger_{\alpha\vec v}b_{\alpha\vec v}=1
\end{equation}
\begin{equation}
b_{\alpha\vec v}b^\dagger_{\alpha\vec v}=0
\end{equation}

	Now, we operate$ b_{\alpha\vec v}$  on the occupation number state$ |1_{\alpha,\vec v}>$corresponding to one particle in the state 
($\alpha,\vec v$)
\begin{equation}
b_{\alpha\vec v}\mid  1_{\alpha,\vec v}> =b_{\alpha\vec v}b^\dagger_{\alpha\vec v}b_{\alpha\vec v} \mid 1_{\alpha,\vec v}>=0
\end{equation}

From equations (18) and (19). Also,
\begin{equation}
b^\dagger_{\alpha\vec v}\mid  1_{\alpha,\vec v}> =b^\dagger_{\alpha\vec v}b^\dagger_{\alpha\vec v}b_{\alpha\vec v} \mid 1_{\alpha,\vec v}>=0
\end{equation}from equation (10). Thus fermion occupation number cannot be changed by operating $b_{\alpha\vec v}$ or $b^\dagger_{\alpha\vec v}$  on 
occupation number state, it remains equal to unity always (eqn.(18)). Thus b's and $b^\dagger$'s lose their meanings as fermion 
annihilation and creation operators.

	How can we then interpret vacuum? The discussion presented above provides a clue to the answer. It is occupied by zeroenergy, which 
for finite $\vec  v$ corresponds to massless fermions. Thus for each ($\alpha, \vec v$) the fermion can appear in either of two mass 
eigenstates with eigenvalues 0 and m, if the fermion is detected only with the mass m.

	If we switch on external interaction these two states get mixed. So the resultant state is not a mass eigenstate in 
general and fermions can make transitions from one state to another. Since occupation number is not fixed the above argument cannot be 
extended to bosons.

	From equations (12) and (13), the free Hamilttonian (eqn.(4)) can be written as

\begin{equation}
	H_0 = \sum_{\alpha,\vec v} (\frac{1}{2} \omega^2_{\vec v} Q^2_{\alpha,\vec v}
	+\frac{1}{2} P^2_{\alpha,\vec v})
\end{equation}

	Which corresponds to quantum mechanical harmonic oscillators with wave functions 
$U_0(Q_{\alpha\vec v})$, $U_1(Q_{\alpha\vec v})$ for the ground and first excited states with energies 0 and 
$\omega_{\vec v}$. Since occupation number is fixed at unity, these wave functions correspond to the mass 0 and m of the fermion. 
We neglect $U_2, U_3 ....$, because they correspond to the mass 2m, 3m etc. of the fermion.

	To see that the above considerations do not contradict established experimental results, we take the simplest example of electrons (fermion) scattered by an external Coulomb potential,

\begin{eqnarray}
A^0&=&\phi \nonumber \\ &=&\frac{1}{\mid\vec x\mid }\nonumber \\ 
&=&\int d^3\vec q\frac{1}{\vec q^2}e^{-i\vec q.\vec x}
\end{eqnarray}

	The interaction Hamiltonian, neglecting anti particles 
\begin{eqnarray}
H^\prime &=& \int\bar\psi\gamma.A \psi d^3\vec x \nonumber \\
&=& \int d^3\vec x \psi^\dagger\psi A^0 \nonumber \\
&=& \sum_{\alpha,\alpha',\vec v, \vec v'} b^\dagger_{\alpha\vec v} 
b_{\alpha'\vec v'} u_{\vec v}^{\dagger (\alpha)} u_{\vec v'}^{(\alpha')}
e^{i(\omega_{\vec v}- \omega_{\vec v'})t}/(m\vec v-m\vec v')^2\nonumber \\
&\sim & \sum_{\alpha,\alpha',\vec v, \vec v'}[Q_{\alpha\vec v}Q_{\alpha'\vec v'}
-iQ_{\alpha\vec v}P_{\alpha'\vec v'}]u_{\vec v}^{\dagger (\alpha)} u_{\vec v'}^{(\alpha')} e^{i(\omega_{\vec v}-\omega_{\vec v'})t}
 /(\vec v-\vec v')^2
\end{eqnarray}

We assume that the interaction is switched on and off at $t =-\infty$  and $+\infty$ respectively. The 
scattering amplitude 
\begin{equation}
C_{fi}=\int_{-\infty} ^\infty H'_{fi}e^{i\omega_{fi}t}dt
\end{equation}

to the lowest order of time dependent perturbation. Since the electron comes from and go to infinity, the energy difference between the initial and final states $\omega_{fi} = 0$.

%\begin{spacing}{1.24}
Let, ($\alpha_1$,$\vec v_1$)  and ($\alpha_2$,$\vec v_2$) be the spin and velocities of the incident and scattered electron. The entire process can be viewed as the mass of the electrons in the states ($\alpha_1,\vec v_1$)and($\alpha_2,\vec v_2$ ) going from m to 0 and vice versa simultaneously. Thus, the initial and the final wave functions will be products of the individual free particle wave functions $ U_1(Q_{\alpha_1,\vec v_1})U_0(Q_{\alpha_2,\vec v_2})$ and $ U_1(Q_{\alpha_2,\vec v_2} )U_0(Q_{\alpha_1,\vec v_1})$ respectively. [in our approach, if an electron (fermion) is attached to the state ($\alpha_1,\vec v_1$), it remains so for ever and under no circumstances can switch to the state($\alpha_2,\vec v_2$). Thus the question of antisymmetrisation of the wave function with respect to the interchange of the two electrons belonging to the states ($\alpha_1,\vec v_1$) and ($\alpha_2,\vec v_2$) doesn't arise.] so,
%\end{spacing}

\begin{eqnarray}
H'_{fi}\sim\int dQ_{\alpha_1\vec v_1}dQ_{\alpha_2\vec v_2}U_1(Q_{\alpha_1\vec v_1})U_0(Q_{\alpha_2\vec v_2}) \nonumber\\
\sum_{\alpha,\alpha',\vec v,\vec v'}[Q_{\alpha\vec v}Q_{\alpha'\vec v'}-iQ_{\alpha\vec v}P_{\alpha'\vec v'}]\nonumber\\
U_1(Q_{\alpha_2\vec v_2})U_0(Q_{\alpha_1\vec v_1})
u_{\vec v}^{\dagger (\alpha)}u_{\vec v'}^{(\alpha')}\nonumber\\
e^{i(\omega_{\vec v}-\omega_{\vec v'})t}/(\vec v-\vec v')^2.
\end{eqnarray}
Only the terms ($\alpha$,$\vec v$) $\equiv$ ($\alpha_1$,$\vec v_1$),($\alpha'$,$\vec v'$) $\equiv$ ($\alpha_2$,$\vec v_2$) and vice versa ,contribute in the sum.Putting $\int dQU^*_1(Q)QU_0(Q)=\int dQU^*_0(Q)QU_1(Q)\sim1$,we get a contribution

\begin{equation}
\sim u_{\vec v_1}^{\dagger(\alpha_1)}u_{\vec v_2}^{(\alpha_2)}e^{i(\omega_{\vec v_1}-\omega_{\vec v_2})t}/(\vec v_1-\vec v_2)^2 + 1 \leftrightarrow 2
\end{equation}

from the first term in the expression for $ H'_{fi}$ . Remembering $P=-i\partial/\partial Q$
 , we get a similar contribution from the second term. Thus
\begin{equation}
C_{fi}=\int_{-\infty}^\infty H'_{fi}dt\nonumber
\sim u_{\vec v_1}^{\dagger(\alpha_1)} u_{\vec v_2}^{(\alpha_2)}\delta(\omega_{\vec v_1}-\omega_{\vec v_2})/(\vec v_1-\vec v_2)^2 + 1 \leftrightarrow 2
\end{equation}

In the non relativistic limit $\vec v_1\approx \vec v_2\approx 0$ and $ u_{\vec 0}^{\dagger(\alpha_1)}u_{\vec 0}^{(\alpha_2)}=\delta_{\alpha_1\alpha_2}$ [from  equations (8)and(9)]. So ,
\begin{equation}
C_{fi}\sim \delta_{\alpha_1\alpha_2}\delta(\omega_{\vec v_1}-\omega_{\vec v_2})/(1-cos\theta)
\end{equation}

           Where $\theta$  is the scattering angle. Thus the scattering cross section
\begin{equation}
\mid C_{fi}\mid^2 \sim cosec^4\theta/2,
\end{equation}

	which is the Rutherford scattering formula.

	Equation (18) and (19) need some clarification. Since $b^2_{\alpha\vec v}=0$  (from equation(10)), $ b_{\alpha\vec v}$ has no inverse. So equations (18) and (19) and for that matter (12) and (13) can't be strong. We would rather take the occupation number state expectation values

\begin{equation}
<b^\dagger_{\alpha\vec v}b_{\alpha\vec v}>=1
\end{equation}
\begin{equation}
<b_{\alpha\vec v}b^\dagger_{\alpha\vec v}>=0
\end{equation}

	Though $b_{\alpha\vec v}$ and $b^\dagger_{\alpha\vec v}$ lose their meaning as fermion annihilation and creation operators, there exist the lowering and raising operators  $a_{\alpha\vec v}$ and $ a^\dagger_{\alpha\vec v}$  for the quantum mechanical harmonic oscillators represented by the Hamilltonian of eqn. (22). So ,
\begin{equation}
Q_{\alpha\vec v}= i( a_{\alpha\vec v}-a^\dagger_{\alpha\vec v})/(2\omega_{\vec v})^{1/2}
\end{equation}
\begin{equation}
P_{\alpha\vec v}=(a_{\alpha\vec v}+a^\dagger_{\alpha\vec v})(\omega_{\vec v}/2)^{1/2}
\end{equation}

With  
\begin{equation}
a_{\alpha\vec v}U_n(Q_{\alpha\vec v})=n^{1/2}U_{n-1}(Q_{\alpha\vec v})
\end{equation}
and
\begin{equation}
a^\dagger_{\alpha\vec v}U_{n-1}(Q_{\alpha\vec v})=n^{1/2}U_n(Q_{\alpha\vec v})
\end{equation}

Corresponding to the decrease or increase of the mass of the fermion by steps of m.

	One may be tempted to identify $ a_{\alpha\vec v}$ with $ b^\dagger_{\alpha\vec v}$ by comparing equations (12), (13), (33), (34). But such a comparison is not feasible in the face of the weak nature of the equations and will lead to inconsistencies.

	To conclude, since each fermionic state is always occupied by a single fermion, there will be no fermion number violation in any process whatsoever. External interactions may however cause a fermion with a finite velocity to go from a massive to the zeromass (energy) state, identified with the fermionic vacuum and viceversa, resulting in apparent annihilation or creation of fermions. The vacuum is however undefined for fermions moving with the speed of light. The very detection of a fermion (eg. neutrino) is an indicator that it is massive. A massless fermion moving with the speed of light can't interact with the detector, because interactions involve intermediate vacuum states, which are undefined in this case.

\textbf{Acknowledgement} : I thank S. Biswas and S.Chakravarty for their unstinting help in the preparation of the manuscript.

\textbf{Reference}:  [1] J. S. Bhattacharyya, S.Biswas, grqc/0211097        

\end{document}